\documentclass[]{spie}  

\usepackage{amsmath,amsfonts,amssymb}
\usepackage{graphicx}
\usepackage{upgreek}
\usepackage[colorlinks=true, allcolors=blue]{hyperref}

\title{The final design of the iLocater spectrograph: An optimized architecture for diffraction-limited EPRV instruments}

\author[a]{Jonathan Crass}
\author[b]{David Aikens}
\author[c]{Joaquin Mason}
\author[d]{David King}
\author[a]{Justin R. Crepp}
\author[a]{Andrew Bechter}
\author[a]{Eric Bechter}
\author[e]{Mahsa Farsad}
\author[f]{Christian Schwab}
\author[a]{Michael VanSickle}
\affil[a]{Department of Physics \& Astronomy, 225 Nieuwland Science Hall, Notre Dame, IN 46556, USA}
\affil[b]{Savvy Optics Corp., 35 Gilbert Hill Rd, Chester, CT 06412, USA}
\affil[c]{Fathom Imaging Systems, 30 Nashua Street, Woburn, MA, 01801, USA}
\affil[d]{Institute of Astronomy, University of Cambridge, Madingley Road, Cambridge CB3 0HA, UK}
\affil[e]{KLA Corporation, 3 Technology Dr, Milpitas, CA 95035, USA}
\affil[f]{MQAAAstro Centre, Macquarie University, NSW 2109, Australia}

\authorinfo{Send correspondence to Jonathan Crass: j.crass@nd.edu}

\begin{document} 
\maketitle

\begin{abstract}
iLocater is a near-infrared, extremely precise radial velocity (EPRV) spectrograph under construction for the dual 8.4\,m diameter Large Binocular Telescope (LBT). The instrument will undertake precision radial velocity studies of Earth-like planets orbiting low-mass stars. Operating in the diffraction-limited regime, iLocater uses adaptive optics to efficiently inject starlight directly into single-mode fibers that illuminate a high spectral resolution (R=190,500 median), cryogenic, diffraction-limited spectrograph. To maximize performance, the spectrograph uses a new design strategy for EPRV instruments, combining intrinsically stable materials for its optomechanical fabrication with precision optical fabrication. This novel combination will enable unique EPRV capabilities for exoplanet and astrophysics studies of the solar neighborhood.

We present the final optical and mechanical designs of the spectrograph system. Ensuring the as-built spectrograph achieves its designed spectral resolution and diffraction-limited performance has required careful control of the end-to-end system wavefront error (WFE) budget. We discuss the efforts undertaken to achieve this goal including minimizing residual WFE in the optical design, assessing diffraction grating WFE performance, optimizing material choices, and requiring precision optical design and fabrication. Our goal is to deliver diffraction-limited performance across the full spectral format, which, combined with intrinsic thermal stability requirements for EPRV science, has driven the selection of silicon optics and Invar optomechanics. The system performance is further optimized using precision (sub-mK) thermal control. This set of design features will allow iLocater to achieve sub-m/s radial velocity precision in the near-infrared, and to serve as the first optimized diffraction-limited spectrograph for EPRV science.
\end{abstract}


\section{Introduction}
\label{sec:intro}  

The discovery and characterization of terrestrial exoplanets has been highlighted as a priority goal in recent astronomy strategic reports. The  \emph{“Worlds and Suns in Context”} science theme of the Astro2020 Decadal Survey made clear that achieving this goal requires the Doppler radial velocity (RV) technique to provide mass, density, and orbital information of exoplanets \cite{NAP26141}. To detect Earth-like planets in Solar analog systems, instrumental precisions at the $\sigma_{\mathrm{RV}}=1-10\,\mathrm{cm/s}$ level are required; however, this precision is beyond even the most cutting edge extremely precise radial velocity (EPRV) instruments \cite{2014Sci...346..809C, 2015arXiv150301770P}. Furthermore, the Doppler variability of exoplanet host stars exceeds the signature of orbiting planets, requiring new instruments and techniques if we are to detect small amplitude rocky worlds \cite{2021arXiv210714291C}. Developing instruments that enable these precision measurements, and support the development of new techniques to mitigate the effects of stellar variability, is an important step to unlocking the full potential for exoplanet detection. 

The iLocater spectrograph is designed to begin addressing both of these challenges. Under development for the Large Binocular Telescope (LBT), AZ, USA, the instrument is one of the first EPRV spectrographs designed to use single-mode fibers (SMFs) fed using adaptive optics (AO) for exoplanet studies \cite{2016SPIE.9908E..19C}. The first of a pair of fiber injection systems for the instrument was successfully commissioned at the LBT in 2019 and is used to feed light into the instrument spectrograph \cite{2021MNRAS.501.2250C}. The spectrograph operates in the near-infrared (NIR) ($\lambda = 0.97-1.31\,\mathrm{\upmu m}$), delivering a median spectral resolution of R=190,500 with diffraction-limited performance being recorded on an H4RG-10 detector. The instrument operating temperature is $T = 80-100\,\mathrm{K}$, with the exact temperature being optimized during operation. Wavelength calibration of the system is provided by a Fabry–Pérot etalon, similar to those used by Maroon-X and NEID \cite{2016SPIE.9912E..29S, 2018SPIE10702E..72S}. The overall design has been developed from fundamental EPRV science requirements, enabling studies to differentiate between exoplanet and stellar signatures while searching for planets around late-type stars. In addition, the instrument mechanical and optical design has been developed to minimize instrument systematics with current technologies.

We present an overview of spectrograph requirements and performance drivers (Section \ref{sec:requirements}) including a series of trade-studies completed as part of the spectrograph design optimization process. The finalized design of the spectrograph is presented in Section \ref{sec:finaldesign} with specific considerations for fabrication being discussed in Section \ref{sec:fabrication}. The program status and current performance is presented in Section \ref{sec:status}.

\section{Instrument Requirements \& Performance Drivers}
\label{sec:requirements}

The iLocater spectrograph design is driven by the need to record high-resolution, high signal-to-noise ratio (SNR) spectra in a stable instrument environment to enable EPRV capabilities. As the instrument is being developed for the LBT, which is equipped with an advanced AO system, iLocater’s design moves away from traditional seeing-limited RV instrument designs to one that operates in the diffraction-limited regime. By using an AO-corrected, diffraction-limited input beam from the telescope, iLocater uses small ($\sim6\mathrm{\upmu m}$ diameter core) SMFs for fiber injection and spectrograph illumination rather than larger multi-mode fibers used by current EPRV instruments. This regime provides numerous benefits including: mitigation of modal noise; enabling high-resolution spectroscopic capabilities to be delivered in a compact mechanical volume; allowing affordable fabrication of optomechanical systems from intrinsically stable materials; and offering a telescope independent spectrograph design that only requires changes in the fiber injection system to be used at other observing facilities \cite{schwab_leon-saval_betters_bland-hawthorn_mahadevan_2012, 2014Sci...346..809C, 2016PASP..128l1001J}.

Development and optimization of iLocater’s performance has focused on four major areas, spanning both the spectrograph directly and broader instrument modules: 

\begin{enumerate}
    \item Maximizing end-to-end system throughput, in particular fiber injection performance, is key to minimizing photon noise effects. The injection of light into SMFs is a potential major area for flux loss and is driven primarily by AO system performance. By selecting a spectrograph bandpass with a high-Strehl ratio and excellent AO performance, fiber throughput can be maximized. In addition, the use of optimized coatings for optics can minimize other throughput losses. 
    \item Designing a spectrograph that achieves the requisite spectral resolution, pixel sampling, and SNR to enable studies, and potential mitigation, of stellar variability. Studies of stellar variability and their impact on RV data is an active area of research; however, delivering data with properties able to address this challenge requires constraints on the spectrograph optical design and overall instrument throughput.
    \item The wavelength coverage of the spectrograph must be optimized to use clean atmospheric transmission windows with limited telluric contamination. 
    \item The spectrograph should be optimized for mechanical and thermal stability to minimize instrument systematic effects that can impact RV precision.
\end{enumerate}

The first three criteria are strongly linked to the overall system optical design and wavelength coverage. To aid with optimization, end-to-end instrument simulations (from host star to spectrograph detector) were used to develop an overall error budget for the instrument performance \cite{2018SPIE10702E..6TB}. This process has allowed the impact of high-level design decisions to be evaluated quantitatively and has enabled an optimized instrument design to be achieved \cite{2019PASP..131b4504B, 2019JATIS...5c8004B, 2020PASP..132i5001B, 2021JATIS...7c5008B}. 

\subsection{Bandpass Selection}
\label{sec:bandpass}

Selection of the spectrograph bandpass had several important considerations. Photon noise is impacted significantly by fiber injection efficiency, which is driven by AO system performance \cite{2020PASP..132a5001B, 2021MNRAS.501.2250C}. As AO correction correlates with the flux being received at the system wavefront sensors (WFS), it is critical to provide as much visible light  as possible to the WFS. Additionally, as AO performance improves with increasing wavelengths, so does fiber injection efficiency. Together, this drives the optimum bandpass for iLocater to the NIR where excellent AO performance can be achieved for fainter stars. Studies of mid- to late-type stars are well suited to this band due to their increased emission and larger exoplanet RV semi-amplitudes compared to solar analogs \cite{2014SPIE.9147E..1GM, 2018SPIE10702E..0WQ}. The Y-band was selected as the primary science band for iLocater given these constraints and also due to its clean transmission window in the NIR \cite{2002PASP..114..708H}.

\subsection{Optical Design Considerations}
\label{sec:opticalconstraints}

The detailed optical specifications for the iLocater spectrograph are driven by the ability to measure, and therefore potentially correct, the effects of stellar variability. Stellar effects such as spots, plague, faculae, and granulation create a measurable skew in the profile of individual stellar absorption lines \cite{2013ApJ...763...95C, 2018ApJ...866...55C, 2019ApJ...879...55C}. Previous work has highlighted the need for high resolution ($R>150,000$) and high-SNR ($\mathrm{SNR} > 300$) RV measurements to be able to resolve these line asymmetries, which, if detected, will provide pathways to their removal and consequently, the detection of exoplanets with RV signatures smaller than those of stellar effects \cite{2021arXiv210714291C}. We have adopted these optical and signal requirements as a baseline for the iLocater spectrograph.

SMFs output a spatially stable Gaussian beam profile and when used to illuminate a diffraction-limited optical system, such as the iLocater spectrograph, particular attention must be given to ensure the beam profile is maintained. Previous studies have shown that, ideally, optics should be sized to fully capture the $2/e^{2}$ diameter of the optical beam to avoid beam truncation that leads to a broadening of the point spread function (PSF)\cite{2012SPIE.8446E..23R}. This broadening would degrade optical performance and, consequently, impact the spectral resolution that can be achieved by a spectrograph. Meeting the $2/e^{2}$ requirement was adopted as the baseline of the iLocater design, however, deviations have been permitted where mechanical constraints preclude this sizing to be achieved. In this situation, a $1.75/e^{2}$  has been adopted which has shown to have a minimal effect on performance. 

Maintaining the instrument beam profile of a diffraction-limited instrument through the entire end-to-end system requires careful management of the wavefront error (WFE) of individual optical components. In the context of EPRV instruments, specific optical aberrations have a different impact on RV precision and residual errors \cite{2021JATIS...7c5008B}. With this in mind, the spectrograph optical design has been carefully optimized to reduce the effects of asymmetric aberrations (for example coma and trefoil) and their potential impact on RV performance. This effort has primarily focused on optimizing the optical design to reduce design residuals that lead to these specific aberrations.

An extensive trade study was completed to identify diffraction-gratings capable of delivering the spectral resolution required to characterize stellar variability. Studies included assessing R4, R6 and R8 gratings with close performance to those that are commercially available as a risk mitigation measure. While an R6 grating was the preferred choice, several effects on performance were noted on spectral resolution, in particular the effects of rotating the grating (gamma angle) to separate the incident and reflected beams. This led to a broadening of the PSF in the dispersion direction and consequently, caused a decrease in spectral resolution due to the ellipticity of the beam onto the echelle, causing a slight anamorphic magnification of the PSF. To limit this effect, the gamma angle of the grating in the system design was minimized as much as possible given optomechanical constraints.

\subsection{Material Selection}
\label{sec:material}

The material used for fabrication of the spectrograph optics and optomechanics is of key importance when considering instrument stability and operation at cryogenic temperatures. Extensive work was done to study the use of different materials in fabrication including an all aluminum (optics and optomechanics) system, and systems with Invar optomechanics and either Zerodur or silicon optics. Three major factors were considered: maintaining thermal stability to enable EPRV science, achieving alignment at cryogenic temperatures, and enabling end-to-end diffraction-limited performance.

EPRV spectrographs must minimize optical path changes arising from temperature variations. The effect can be incredibly small, requiring milli-Kelvin levels of thermal stability, however, optomechanical design choices can help loosen thermal stability requirements. For example, to first order, single material systems (where the optics and optomechanics are fabricated from the same material) self-compensate for temperature changes as the radius of curvature of optics compensates for global position changes on an optical bench. Alternatively, by using materials with a low coefficient of thermal expansion (CTE) at the instrument operating temperature, a system which is intrinsically stable can be fabricated, minimizing thermal expansion. However, to do this in a multi-material system requires a careful match of CTE and tuning of the operating temperature to allow the instrument to behave as if it was a single material. While an Invar and Zerodur instrument may initially seem advantageous due to these two materials having low CTE at room temperature, at cryogenic temperatures, the CTE mismatch between the two can be challenging to combine when seeking to achieve stability and an aligned optical system. An Invar and silicon instrument however is well suited to behave as a single-material system around 100\,K, providing an optimized environment for EPRV science \cite{Crass_2022}.

Given iLocater’s need to maintain diffraction-limited performance, the surface figure of delivered optics was studied, both in terms of wavefront performance as well as scattered light. The component figure error was deemed to be higher for aluminum than glass optics due to fabrication processes, increasing wavefront error in the delivered iLocater system. When combined with considerations of surface quality and their impact on scattered light, glass optics were preferred. For these reasons, iLocater uses Invar optomechanics combined with silicon optics. 

\section{Final Instrument Design}
\label{sec:finaldesign}

\begin{figure} [ht]
   \begin{center}
   \begin{tabular}{c} 
   \includegraphics[width=\textwidth]{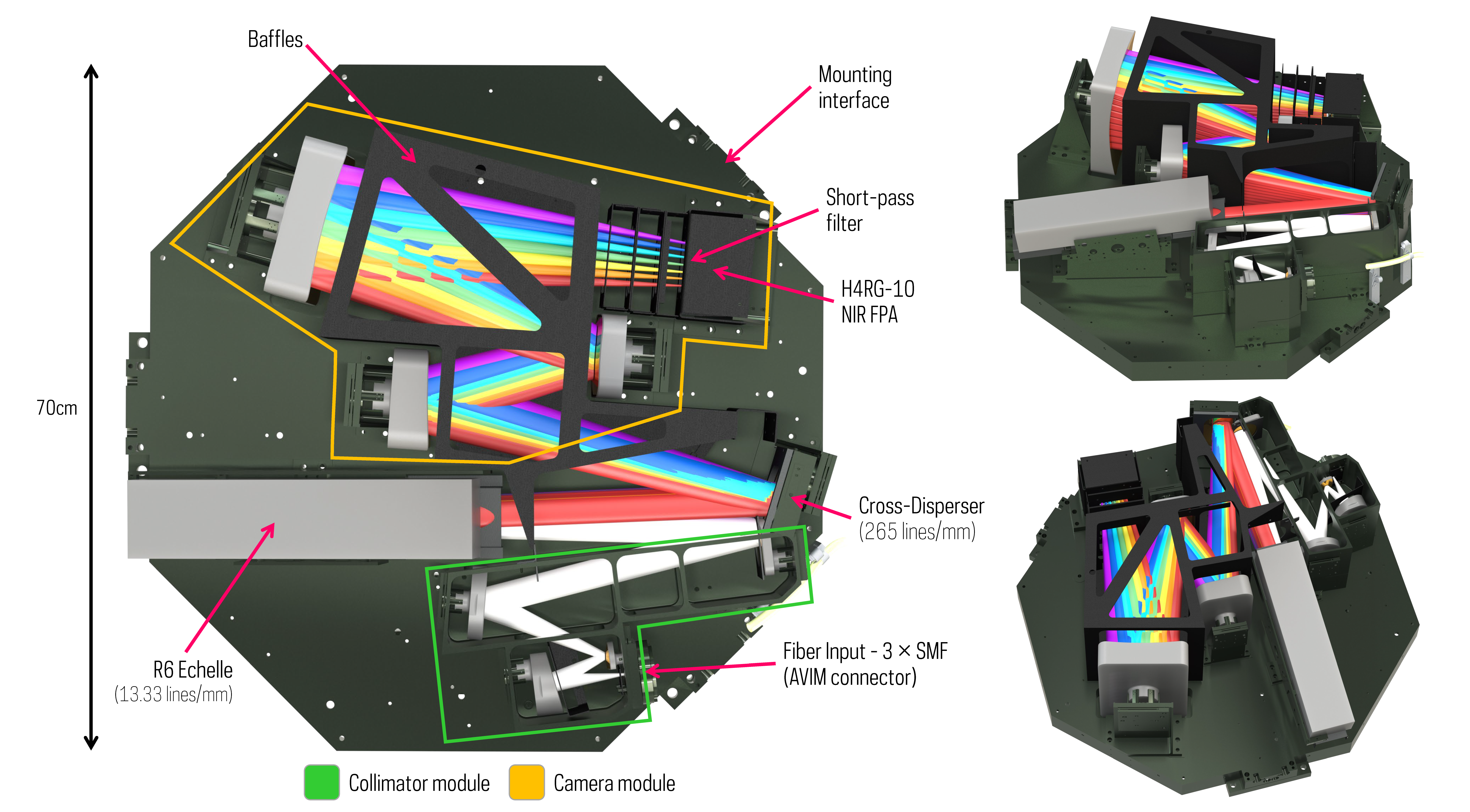}
   \end{tabular}
   \end{center}
   \caption[CAD rendering of the spectrograph] 
   { \label{fig:spectrograph} 
Left: An annotated top-down CAD rendering of iLocater spectrograph showing the collimator and camera modules, and key components. Right: side view renderings of the spectrograph.}
   \end{figure}

The final design of the spectrograph (Figure \ref{fig:spectrograph}) combines the outcomes of all the trade-studies completed as part of the iLocater program to deliver an instrument capable of achieving EPRV science. The system comprises two independent optical modules linked by a pair of diffraction-gratings. The collimator module takes the output beams of three closed-packed SMFs (SM980) and collimates them onto the R6 echelle grating (13.3 lines/mm). The beam is then cross-dispersed using a second grating (265 lines/mm) before being focused using the camera module onto an H4RG-10 detector, which is controlled by an ARC imaging array controller. The camera module serves as the main base-plate for the entire system with the collimator module being situated and pinned into position on the camera module. The camera base-plate is 72.5\,cm in diameter and has been lightweighted to reduce overall thermal mass and to aid with instrument cooldown. Unlike seeing-limited spectrograph designs, the small input used for spectrograph illumination needs to be magnified onto the detector which generates a fast optical system in the collimator module (f/4.54) and a slow camera system (f/16.2). Significant work has gone into ensuring the collimator system meets its stringent optical tolerances.

All optics in the spectrograph are gold-coated silicon to minimize potential RV error sources arising from chromatic effects in transmissive components. A high-reflectivity enhanced gold coating has been applied to maximize overall throughput and minimize scattered light effects, achieving an end-to-end system throughput (without grating contributions) of 88.5\%. Z306 painted baffles are installed throughout the system to reduce stray light and limit the unwanted illumination of optical surfaces. The overall spectrograph is housed within a Z306 coated radiation shield to further limit stray light effects.

\subsection{Optical Performance}
\label{sec:opticalperformance}

The spectrograph optical design delivers high-resolution spectra for 38 orders ($m=114-152$) spanning $\lambda = 0.97-1.31\,\mathrm{\upmu m}$. The three fibers used for illumination (one for each LBT primary mirror and one for calibration) generate three independent traces per order. The fibers are clocked to provide adequate space between traces while maximizing the number of orders on the detector. The spectral resolution of the system has a median resolution of R=190,500, however, the use of an R6 grating causes a change in dispersion across each order and between orders. The spectral resolution ranges from a minimum of 125,000 to a maximum of 281,000 with pixel sampling per resolution element ranging from 2.2 to 3.1 with a median of 2.7. The wavefront error of the designed end-to-end system is $\lambda/10$ at 632.8nm without diffraction-gratings, delivering diffraction-limited performance.

\subsection{Mount Designs}
\label{sec:mountdesigns}

\begin{figure} [ht]
   \begin{center}
   \begin{tabular}{c} 
   \includegraphics[width=0.75\textwidth,trim={9cm 0 0 0},clip]{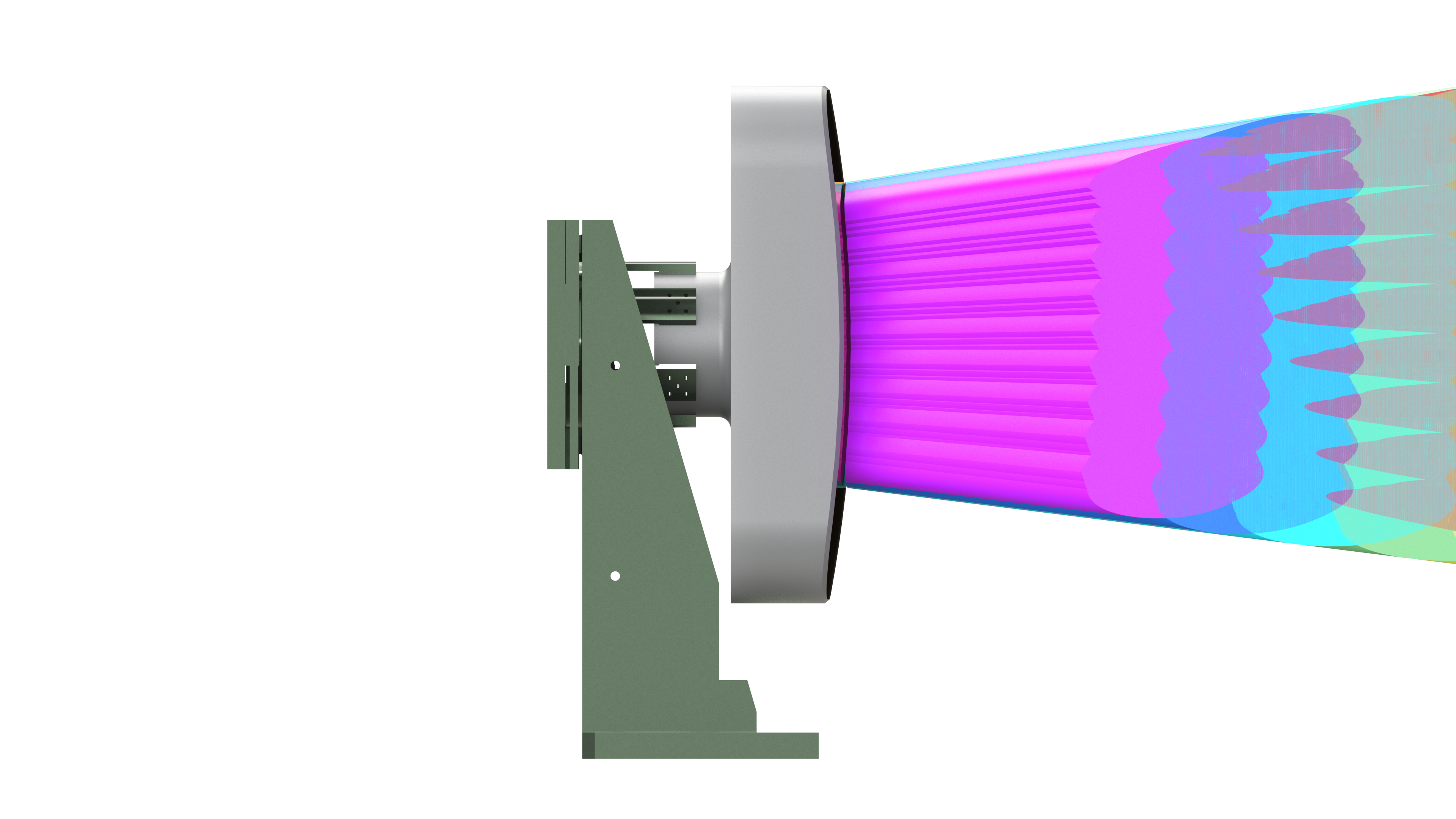}
   \end{tabular}
   \end{center}
   \caption[CAD rendering of the mount design] 
   { \label{fig:mount} 
CAD rendering of one of the optomechanical mounts in the camera module. The silicon optic includes a rear post which is bonded to Invar flexures. The flexure assembly is then secured to a further supporting structure.}
   \end{figure}

The optomechanical mount design focuses on maximizing thermal stability of the optical system while minimizing distortions on the optical surfaces, particularly when considering thermal cooldown effects. A rear ‘post’ design has been used for the mounts (Figure \ref{fig:mount}) where a set of  Invar flexures are bonded to a rear post machined into each optic. This type of design limits the strain in the optic and ensures no distortion from the mount interfaces reaches the front surface causing wavefront errors. The flexures then mount to a plate and shim assembly for adjustability.  The collimator optics are mounted to a sub-structure Invar box for the entire module assembly, while the camera uses pedestal mounts secured to the main integration base. At final assembly, all bolted interfaces are pinned for robust environmental optical stability.  

The diffraction-gratings in many EPRV instruments drive critical thermal constraints given the tight tolerances on line spacing and global movement. Within iLocater, the grating mount design constrains the gratings using spring flexures to ensure the front-surface of the grating serves as the reference interface for positioning. Fine adjustment of the cross disperser follows the technique for the camera mirrors using shimming, while the echelle grating uses Invar rollers to adjust global tip/tilt of the grating front surface. This overall design minimizes distortion on the reflective grating surface by only providing lateral constraints while adequately supporting the grating to minimize surface distortions caused by gravitational sag.

Specific consideration has been given to the detector and fiber mount to ensure stability at these key elements. Both systems are supported on Invar optomechanics with athermalization elements being incorporated into the design to aid with focus stability between room temperature and the instrument operating temperature. The instrument fibers are housed within an AVIM-type connector, which has precision features machined into the ferrule to allow repeatable positioning, and pre-loading to ensure stability. The detector mount uses Invar to maintain the metering path to both the detector interface plate as well as a short-pass filter which is situated just prior to the detector focal plane. Despite the close match in thermal expansion to the operational temperature between the Invar structure and the silicon optics, athermalization compensators are desired to maintain the precision focus without requiring cryogenic adjustment.  The fiber mount design contains a compensation spacer to maintain focus from room temperature to the instrument operating temperature; an axial flexure preloaded with an aluminum spacer  provides the expected focus compensation for the entire Invar and silicon system.

\section{Instrument Fabrication}
\label{sec:fabrication}

Components within the spectrograph have undergone a careful fabrication process to ensure overall performance and minimize risks during system integration. This process has included appropriate machining, heat and cool treatments of the Invar hardware in the optical path, in addition to precise machining and polishing of optics to achieve overall throughput and minimize WFE. Where necessary, test components fabricated from aluminum have been produced prior to the final system fabrication. This has been done to ensure the machining processes used achieved the necessary performance and tolerances for the final instrument.

\subsection{Optomechanics Fabrication}
\label{sec:optomechanicsfab}

The Invar elements used to support and interface with optical components have undergone careful treatment processes to ensure the CTE of the material closely matches the values used in the system design and simulations. The raw Invar was annealed prior to machining to ensure consistent CTE throughout the material. A second re-annealing following the coarse machining was completed to remove any possible CTE variations imparted during the machining process. A precision final machining process ensured compliance with optical and mechanical interfaces and the final components underwent a mild heat treatment for final stress relief and to prematurely age the Invar to minimize long-term mechanical changes. For the collimator module, this was followed by a cryogenic thermal cycling to the instrument operating temperature prior to optics installation to further stabilize residual thermal strains. The same cryogenic cycling was used to verify the performance of the epoxy and representative bond geometry used to attach the flexures to the optics. To prevent long-term degradation, all Invar components have been coated with corrosion protection.

\subsection{Optics Fabrication \& Integration}
\label{sec:opticsfab}

The silicon optics used in the spectrograph have been manufactured in a multi-stage process by Nu-Tek Precision Optical Corp. The front face of the optics has been manufactured using diamond turning that was followed by machining to define the global optic shape, including the rear mounting post. The surface has been finished using a combination of traditional polishing and magnetorheological finishing (MRF) to deliver minimal scatter and low residual WFE. An enhanced gold coating has been applied to all optical components with greater than a 98\% reflection performance across the instrument bandpass. An initial coating of the optics showed a slowly emerging surface degradation over time, impacting throughput and scattered light. The optics were re-coated to mitigate the issue and no further degradation has been noted.

Integration of the camera and collimator optics into the optomechanical mounts has comprised several key steps. First, the optics were bonded to their Invar flexure mounts to allow handling and positioning.  A coordinate measuring machine (CMM) was used to provide coarse positioning of the optical elements within the optomechanical system with fine adjustment being achieved using pushers and reference markers. Once an adequate alignment was achieved so that fringes were observed when illuminating the system with an interferometer, a baseline performance was noted. Aluminum shims were then added to improve stability versus initial adjustable elements, with one optic being adjusted at a time until the previous optical performance had been recovered. The same process was subsequently completed to switch the shims from aluminum to Invar which was followed by a optimization and hybrid metal pin/liquid pinning to achieve final optics stability.

\begin{figure} [ht]
   \begin{center}
   \begin{tabular}{c} 
   \includegraphics[width=0.75\textwidth]{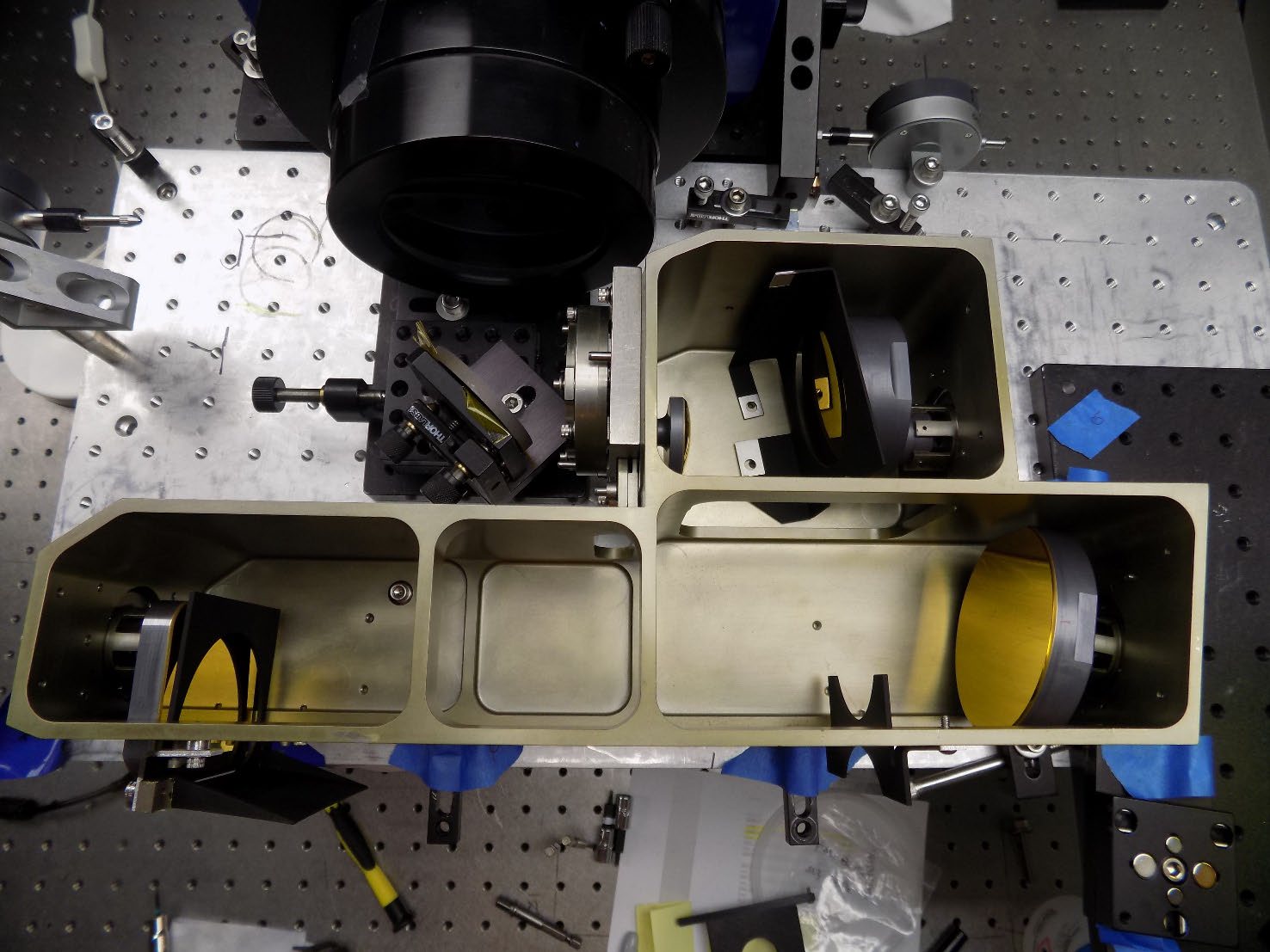}
   \end{tabular}
   \end{center}
   \caption[The iLocater collimator module] 
   { \label{fig:collimator} 
The iLocater collimator module during final end-to-end wavefront testing. Z306 baffles are situated at key locations in the collimator optical train to reduce scattered light effects.}
   \end{figure} 

\begin{figure} [ht]
   \begin{center}
   \begin{tabular}{c} 
   \includegraphics[width=0.95\textwidth]{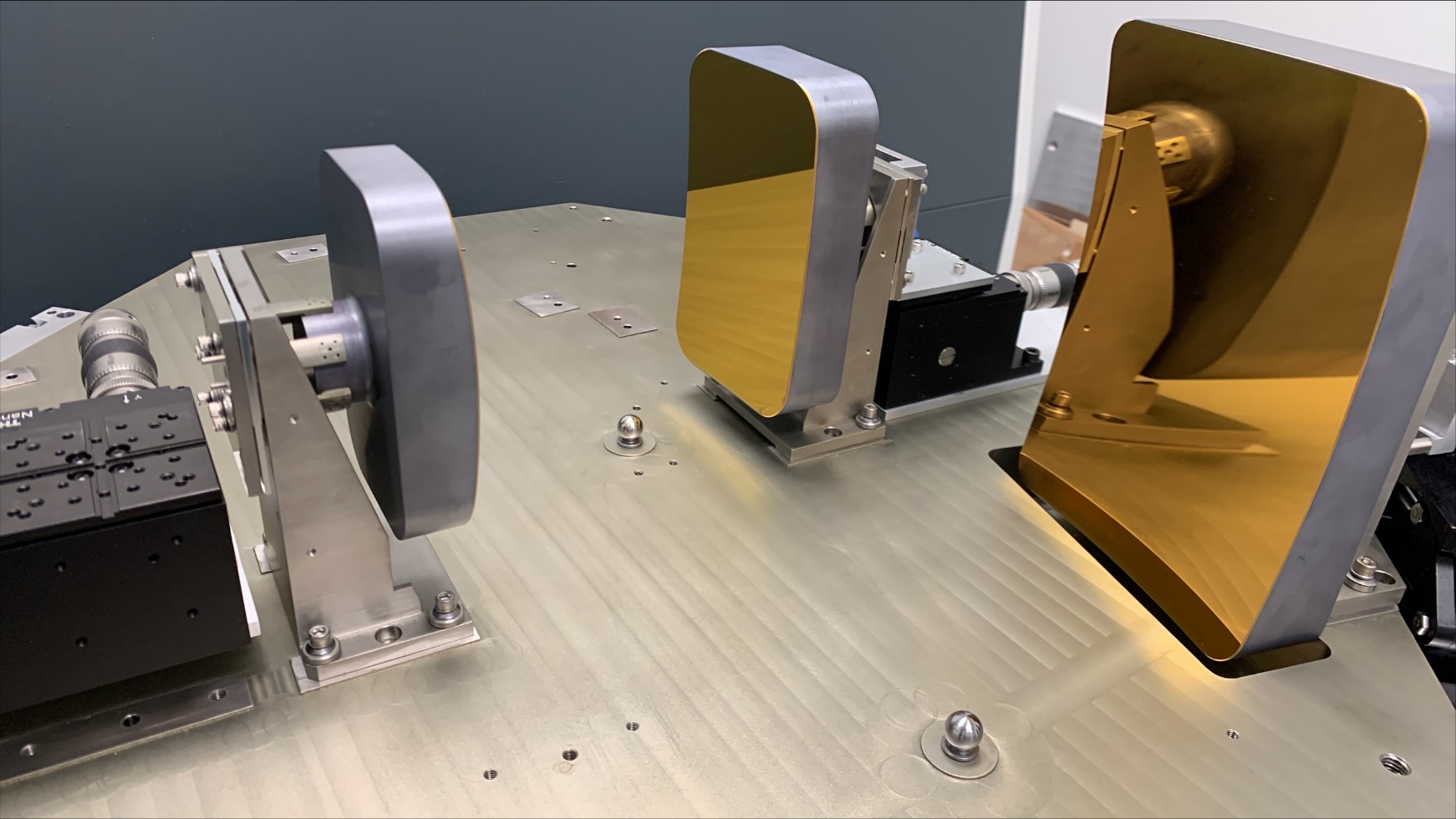}
   \end{tabular}
   \end{center}
   \caption[The iLocater camera module] 
   { \label{fig:camera} 
The iLocater camera module during initial CMM alignment.}
   \end{figure}

The diffraction-gratings within the spectrograph are manufactured from etched silicon \cite{2014SPIE.9151E..1GB}. This choice has primarily been driven by the need for low WFE to achieve diffraction-limited system performance which is beyond the performance of commercially available gratings which have a WFE typically around $\lambda /2 -\lambda / 4$. Precision etched silicon gratings provide low WFE ($<\lambda /20$) and improved diffraction efficiency compared to commercial gratings, increasing overall instrument throughput.

\section{Status \& Performance}
\label{sec:status}

The entire spectrograph system is currently undergoing final integration. The collimator module (Figure \ref{fig:collimator}) has been successfully completed with diffraction-limited performance being achieved across the three illumination fibers (WFE = 0.038, 0.028 and 0.040 waves RMS respectively). The camera module (Figure \ref{fig:camera}) is undergoing final alignment and has achieved a nominal WFE performance of 0.05 waves RMS across the full optical path. Collectively, these modules are expected to meet the WFE required to deliver an end-to-end diffraction-limited optical system.

Integration of the instrument spectrograph and cryostat is expected to continue through 2022 with cryogenic testing of the end-to-end optical system completing optical verification. This testing will initially focus on assessing PSF quality at the detector focal plane using emission lamps and iLocater’s wavelength calibration system. If required, it is possible to undertake interferometric testing of the camera and collimator modules independently at cryogenic temperatures using a dedicated feedthrough port on the instrument vacuum chamber. Optical testing will be followed by a thermal stability assessment of the spectrograph using the iLocater thermal control system \cite{Crass_2022}.

\section{Conclusions}
\label{sec:conclusions}

The iLocater spectrograph has been designed from fundamental EPRV science requirements to deliver an instrument that is highly optimized to undertake precise RV studies of exoplanet systems. The instrument trade studies have highlighted key drivers of instrument performance and have guided instrument optimization and design decisions. By using SMFs for illumination, iLocater operates in the diffraction-limited regime achieving high spectral resolution in a compact instrument. This will enable unique capabilities to study, and potentially mitigate, stellar variability. Additionally, the use of Invar optomechanics combined with silicon optics, creates a diffraction-limited end-to-end system with intrinsic stability that is well suited to EPRV science programs. The spectrograph modules have already demonstrated their performance, and overall system integration is expected to be completed in 2022.

The diffraction-limited nature of the iLocater spectrograph means its design is telescope independent. Therefore, it can be used as a basis for other high-resolution SMF-fed spectrographs on other AO-equipped telescopes, with changes only being required to the front end fiber injection system. By using the design as a baseline for new diffraction-limited instruments, the extensive design and trade studies completed as part of the iLocater program will provide a well optimized spectrograph architecture. This will help to minimize the development timelines and costs of future SMF-fed EPRV instruments.


\acknowledgments 
 
This material is based upon work supported by the National Science Foundation under Grant No.~1654125 and 2108603.

\bibliography{report} 
\bibliographystyle{spiebib} 

\end{document}